\documentclass[11pt,twocolumn]{article}
\usepackage{authblk}
\usepackage{verbatim}
\usepackage{cite}
\usepackage[pdftex]{graphicx}
\usepackage{amsmath}
\usepackage{hyperref}

\makeatletter
\renewcommand\@maketitle{%
\hfill
\centering
\begin{minipage}{\textwidth}
\vskip 1em
\let\footnote\thanks 
{\LARGE \textbf \@title \par }
\vskip 2em
{\large \@author \par}
\end{minipage}
\vskip 2em \par
}
\makeatother

\begin{document}

\title{The Dynamics of Handwriting Improves the Automated Diagnosis of Dysgraphia}

\author[1,\footnote{Contributed equally},\footnote{Corresponding author, \url{konrad.zolna@gmail.com}}]{Konrad~\.Zo\l{}na}
\author[2,$^{*}$]{Thibault~Asselborn}
\author[3,$^{*}$]{Caroline~Jolly}
\author[4]{\\Laurence~Casteran}
\author[4]{Marie-Ange~Nguyen-Morel}
\author[2]{Wafa~Johal}
\author[2]{Pierre~Dillenbourg}

\affil[1]{Jagiellonian University,~GMUM}
\affil[2]{\'{E}cole~Polytechnique~F\'{e}d\'{e}rale~de~Lausanne,~CHILI~lab}
\affil[3]{Universit\'{e}~Grenoble~Alpes,~CNRS,~LPNC~UMR~5105}
\affil[4]{Grenoble~Hospital,~Reference~Center~for~Speech~and~Learning~Disorders}

\maketitle

\begin{abstract}

\small Handwriting  disorder (termed dysgraphia) is a far from a singular problem as nearly 8.6\% of the population in France is considered dysgraphic. Moreover, research highlights the fundamental importance to detect and remediate these handwriting difficulties as soon as possible as they may affect a child's entire life, undermining performance and self-confidence in a wide variety of school activities.\\
At the moment, the detection of handwriting difficulties is performed through a standard test called BHK. This detection, performed by therapists, is laborious because of its high cost and subjectivity.
We present a digital approach to identify and characterize handwriting difficulties via a Recurrent Neural Network model (RNN). The child under investigation is asked to write on a graphics tablet all the letters of the alphabet as well as the ten digits. Once complete, the RNN delivers a diagnosis in a few milliseconds and demonstrates remarkable efficiency as it correctly identifies more than 90\% of children diagnosed as dysgraphic using the BHK test.\\
The main advantage of our tablet-based system is that it captures the dynamic features of writing -- something a human expert, such as a teacher, is unable to do. We show that incorporating the dynamic information available by the use of tablet is highly beneficial to our digital test to discriminate between typically-developing and dysgraphic children.

\end{abstract}

\normalsize

\section{Introduction}\label{sec:introduction}
\label{sec:introductionHandwriting}

Even in the era of digital learning, handwriting is still a critical element of early childhood education. 
Handwriting is a complex task involving cognitive, perceptual, attentional, linguistic, and fine motor skills \cite{bourdin2000graphic,feder2007handwriting, mccutchen2011novice}. In France, children learn to write cursively around the age of 5 (preschool) and complete their handwriting mastery around ten years later \cite{accardo2013development, ziviani2006development}.  During learning, handwriting evolves both at qualitative (letter shapes and legibility) and quantitative levels (speed) \cite{blote1991longitudinal, rueckriegel2008influence, accardo2013development}. In a constant search for efficiency, handwriting progressively becomes automated thanks to motor programs, which allow the motor control system to produce integrated movements. These programs appear around the age of 8, and it is generally admitted that automation is complete around the age of 14 \cite{blote1991longitudinal,chartrel2008impact,feder2007handwriting,rueckriegel2008influence,vinter2010effects,ziviani2006development}. A significant breakthrough during recent decades in the study of handwriting came from the development of graphics tablets and dedicated software, which enables the high-frequency sampling and subsequent analysis of handwriting dynamics. These tools greatly contribute to our current knowledge of handwriting kinematics, the development of motor skills for planning and controlling handwriting movements \cite{pontart2013influence}, and the understanding of neuromotor deficits in handwriting disorders.  
In the CoWriter project \cite{jacq2016building,lemaignan2016learning,johal2016child}, we aim to adapt handwriting training for children with difficulties by using a learning by teaching approach in which children teach a humanoid robot how to write on a graphics tablet.

Despite correct training, between 5\% and 25\% of children never master handwriting like their peers. With the increasing cognitive demand for school work, handwriting rapidly becomes a limiting task for these children who cannot handle simultaneous efforts such as handwriting, grammar, orthography, and composition. These students may then rapidly face more general learning difficulties. Thus, it is crucial to detect and remediate handwriting difficulties as soon as possible \cite{feder2007handwriting,christensen2009critical}.

The handwriting tests are not difficult; children are asked to copy a standard text, which is then evaluated based on a predefined set of features. In French-speaking countries, the “BHK test” (The Concise Assessment Scale for Children’s Handwriting) \cite{hamstra1993longitudinal,charles2004echelle,soppelsa2012evaluation} is widely used for diagnosing dysgraphia. This test is recognized by health insurance companies, which pay the costs of both diagnosis and treatment.

But when it comes to diagnosing dysgraphia with the help of the BHK test, a number of difficulties may arise. These are related to the amount of time required for scoring the tests, variability across evaluators, and most notably the time lag – often a period of 6 months or more – between initial concerns about a child’s handwriting and the opportunity to consult with an expert. 

In this context, we develop a new method to allow rapid and accurate discrimination between typically-developing (TD) and dysgraphic children. The goal of our study is to create a diagnosis tool independent of human expertise to avoid any subjectivity. We use a simple writing task, the alphabet task, which was proven previously to predict children handwriting performance and to discriminate between proficient and non-proficient writers \cite{rosenblum2005using,berninger1991scientific,berninger1998language,graham2001manuscript}. We showed that the main advantage of this tablet-based test it that it exploits the dynamic features of handwriting. Two models were compared: a Convolutional Neural Network only analyzing the data as a static image (like the pen-paper tests used nowadays) with a Recurrent Neural Network using the dynamical aspect of the data. We showed that the latter outperformed the former.

\section{Related work}

\subsection{Handwriting legibility analysis}

Assessing the legibility or readability of handwriting is not a new challenge as studies relating to this topic exist since the beginning of the twentieth century. The first scaling method was developed by Thorndike \cite{thorndike1912handwriting} in 1910. This constituted a very important contribution "not only to the experimental pedagogy but to the entire movement for the scientific study of education" \cite{ayres1912scale}. Thorndike compared his invention to the thermometer. "Just as it was impossible to measure temperature beyond the very hot, hot, warm, cool, etc., of subjective opinion, so it had been impossible to estimate the quality of handwriting except by such vague standards as one's personal opinion that given samples were very bad, bad, very good, etc." \cite{ayres1912scale}.

Until now, two approaches are used to evaluate handwriting. The first is a global holistic method that evaluates the handwriting quality as a whole, while the second measures it according to several predefined criteria \cite{rosenblum2003product}. 

The global holistic approach provides an overall judgment on the quality of handwriting by comparing handwriting samples previously sorted depending on quality. As a typical example of this approach, Ayres developed in \cite{ayres1912scale} a scale in which teachers can mark the legibility of a handwriting sample by comparing it to eight handwriting references of increasing quality included with the scale. Assessment of the writing quality relies on subjective judgment by the teacher. After that, several updated scales with more objectivity were developed \cite{bezzi1962standardized, freeman1959new, herrick1963evaluation}.

The second approach for handwriting legibility analysis is based on predefined criteria (e.g., letter form, letter size, spacing, and line-straightness). The judgment is then made by individually grading these criteria and summing the sub-scores.  Among the large number of tests developed during the past 40 years using this approach, we can cite \textbf{the concise Evaluation Scale for Children's Handwriting (BHK)} test \cite{hamstra1987concise} as the primary reference for the diagnosis of dysgraphia in Latin alphabet-based languages \cite{hamstra1993longitudinal,karlsdottir2002problems,overvelde2011handwriting,charles2004echelle,soppelsa2012evaluation}. We can also cite more recent tests, such as the \textbf{Evaluation Tool of Children's Handwriting (ETCH-C)} \cite{amundson1995evaluation} and the \textbf{Hebrew Handwriting Evaluation (HHE)} \cite{erez1999hebrew}, which take into account other features of handwriting. In the \textbf{ETCH-C} test, for example, the pencil grasp, pencil pressure, and in-hand manipulation are additional criteria taken into account. In the \textbf{HHE}, the examiner is asked to observe the pencil and paper position, the body posture and stabilisation, and the fatigue. However, the observation of all criteria remain partly subjective as a human being makes them.

With the emergence of new tools (e.g., digital tablet), the addition of several variables (hidden so far) to the analysis of handwriting legibility is now possible. In particular, the analysis of dynamic features of handwriting allows a better characterisation of childhood handwriting difficulties \cite{rosenblum2003product}. Several techniques have then been proposed to classify the handwriting legibility, including the dynamics of the process \cite{danna2013signal, di2008dynamic, rosenblum2017identifying}, thus leading to better accuracy and less subjectivity.

\subsection{Models used in handwriting \mbox{analysis}}

Recently, much research has generated models for handwriting analysis using machine learning. Indeed, machine recognition of handwriting is used in various fields, such as reading postal addresses on envelopes, amounts written on bank checks, and signature verification. Models are divided into offline and online recognition. Offline recognition focuses on the image of the handwritten text, while in online recognition, the location of the pen-tip is recorded as a function of time \cite{plamondon2000online}. As the offline/online designation might not be clear, we change the terminology for the remainder of this paper to static/dynamic as we believe these are more relevant.

Static systems are less accurate than dynamic ones due to the absence of temporal data (dynamics of writing), which contains information that may be relevant \cite{plamondon2000online} for the model. Only the image of the handwritten text is available to the model.  MNIST \cite{lecun1998gradient} is the most widely used benchmark for isolated handwritten digit recognition. Many machine learning techniques have been used to classify the MNIST dataset from the multi-layer perceptron (MLP) \cite{simard2003best} to complex variants of support vector machines (SVM) \cite{decoste2002training}. However, CNN models seem to perform best by providing outstanding results on the MNIST dataset as well as on Latin letters and Chinese characters, for example, \cite{ciregan2012multi}.

For dynamic recognition, the temporal information about handwriting is available to the model. Different architectures of models are then used for the classification. For example, a Hidden Markov Model (HMM) \cite{marti2001using} or a hybrid system gathering time-delay neural network and HMM \cite{jaeger2001online} has been used in the past twenty years. Since then, RNNs have been used with great success surpassing all previous results in this field \cite{graves2008unconstrained,graves2009novel}.

While the literature reports a large number of models to recognise digits, characters, words or sentences, very few works aimed to assess handwriting legibility for dysgraphia detection purposes.
Rosenblum et al. \cite{rosenblum2017identifying} used a SVM classifier on a set of manually extracted handwriting features to classify between dysgraphic and non-dysgraphic Hebrew writing children. The SVM binary classification led to the accurate labeling of 89 out of 99 writing products (89.9\% accuracy). In \cite{asselborn2018automated}, Asselborn et al. used a Random Forest classifier to correctly identify 98\% of the 56 dysgraphic children in their database. Their method was taking in input the BHK test (latin alphabet).

With our work, we embrace the same goal and propose to create an easier test to effectively detect dysgraphia in a classroom context. Instead on focusing on the analysis of an handwritten text, task that appears to be challenging for a wide range of children, we developed a test working on individual letters (and numbers) that are taught earlier in school curriculum \cite{piasta2010developing,asselborn2018bringing,ritchey2008building}.
In addition, by characterizing difficulties independently for each letter, we envision to build a personalized training system that would generate words with letters specifically selected to suit the training needs of the child. 

The next section presents our motivations and the machine learning tools used to build our models for the diagnosis.

\section{Motivations and Technical Grounding}\label{sec:neuralNetworks}
The overall goal of this project is to enable teachers to evaluate in a fine grain handwriting in order: (1) to detect children in high difficulties and at risk for dysgraphia, (2) to generate a handwriting profile for the child pointing gyphs that should be practiced more in later-on handwriting training.
For this, we propose to build a system able to assess handwriting legibility automatically.
We present below pros and cons of several methods and argument our choices.

Artificial neural network (ANNs) are particularly useful in the situations where the relationship between inputs and outputs is complex and challenging to identify by a non-expert human observer.
Practice has proven that artificial neural networks may be successfully employed to solve many of practical problems. They have been used in a variety of applications, for example, machine translation \cite{cho2014learning}, image recognition \cite{krizhevsky2012imagenet}, and even playing computer games on the level beyond human skills \cite{mnih2013playing}.

In this work, we focus on two types of ANNs: LSTM recurrent neural networks (RNN) \cite{hochreiter1997long} and convolutional neural network (CNN) \cite{lecun1998gradient}. RNNs were introduced to analyse and interpret sequences of data. Hence, this approach meets our applicative setup where each glyph\footnote{A \textbf{glyph} is a graphic symbol that provides the appearance or form for a character. A glyph can be an alphabetic or numeric. In this paper, we focus on lower case cursive alphabetical [a-z] and numerical [0-9] glyphs.} may be understood as a trajectory or the sequence of consecutive positions of the pen tip on the paper. This approach focuses on the temporal aspect of data and considers the dynamics of writing. This characteristic is known, and RNN has been applied to handwriting tasks before \cite{graves2013generating}. For this reason, we believe that RNN can provide better results to our applicative problem compared to other methods using static images (including the original way of evaluating BHK tests).

CNNs are widely used tools for deep learning. They are known to be particularly well suited for applications with images as inputs, although they are now more frequently used in other applications involving text \cite{kim2014convolutional} or signals \cite{li2010automatic} as inputs. CNNs have become a standard in text recognition giving outstanding results \cite{wang2012end,lecun1998gradient}. 

In addition to the different architectures of the network, it is important to notice that contrary to RNN, CNN takes the 2D image representing the glyph as input, meaning that the dynamic of the movement is lost. We present here a new promising method based on a neural network taking into account dynamic features of handwriting to discriminate between TD and dysgraphic children.

\section{Methods}\label{sec:data}

\subsection{Participants}

The present study was conducted in accordance with the Declaration of Helsinki, was approved by the University of Grenoble Alps' ethics committee (agreement n° 2016-01-05-79). It was conducted with the understanding and written consent of each child's parents and the assent of each child and in accordance with the ethics convention between the academic organisation (LPNC-CNRS) and educational organisations. A total of 971 typically-developing children were recruited in 14 schools from various suburbs to ensure different socio-economic environments (\emph{TD dataset}). 43 classes were included from pre-school to fifth-grade. None of the TD children included in the study presented known learning problems or neuromotor disorders. Twenty-four dysgraphic children recruited at the Learning Disorders Center of Grenoble hospital (Centre R\'{e}f\'{e}rent des Troubles du Langage et des Apprentissages, CHU Grenoble) were also included in the study (\emph{D dataset}) and were all diagnosed as dysgraphic based on their BHK scores.
 
\subsection{Data collection}

Children were asked to write cursively, without a time limit, the 26 letters of the alphabet in lower case, as well as 10 digits, randomly dictated. Two procedures following different dictations were performed, the first in the middle of the school year (January-February) and the second at the end of the school year (May-June). We checked that the dictation order did not affect children performances (data not shown). Dictations were performed on a sheet of paper placed on a Wacom© Intuos 4 A5 USB graphic tablet (sampling frequency = 200 Hz; spatial resolution = 0.25 mm). The sheet of paper was used to place the children in their usual handwriting conditions as asking them to write directly on the tactile surface might be different due to a different friction coefficient \cite{alamargot2015does}. All tracks were monitored using the Scribble software developed in the LPNC laboratory \cite{jolly2014one,jolly2014analysis}.

\subsection{Analysis of handwriting and identification of dysgraphic children}

\subsubsection{Scarcity of annotated data}\label{scarcity}

As mentioned above, handwriting is a crucial skill to acquire and paramount for many school activities. Hence, children are taught handwriting in a standardised way from the beginning of their school education. As a result, the vast majority of data collected is inadequate compared to with what was taught in school (resulting in an over-representation of "good" examples). In other words, the class of glyphs that we are particularly interested in ("bad" examples) are underrepresented. In addition, we have a significant variety of instances that belong to the underrepresented class (illegible glyphs) as compared to a relatively small variance in the overrepresented class--there are only a few ways to write a letter correctly, while there are many ways to do it incorrectly. 

Hence, the discrimination between typically-developed and dysgraphic children seems to be challenging as the class of illegible glyphs we want to describe is underrepresented and varies a lot. We argue that using a black box to solve this problem would require a considerable amount of data, which we find infeasible. We are then left to work in a regime of underrepresented positive cases meaning that a more sophisticated solution must be developed. We applied an idea of transferring learning to address this challenge.

\subsubsection{Transfer learning}
In machine learning, transfer learning is an approach where a new task is solved through the transfer of knowledge from a related proxy problem that is simpler to solve \cite{torrey2009transfer}.

A small group of shapes are fixed to be letters or digits, and people are trained to write and read these symbols. The glyphs written by dysgraphic children tend to be harder to decode as they are not written conventionally. This is why we believe that the problem of identifying dysgraphic children may be resolved by using an approach that mimics humans. Hence, we decided to train a recognising model to classify glyphs and use its prediction to discriminate dysgraphic children. We assume that if the model fails to predict which glyph is the given one, then the writer has a higher probability to be dysgraphic. Our recognising model is trained to discriminate glyphs, in other words, the model assesses legibility of a given glyph. It is important to notice that this approach (using the predictions of the recognising model by a high-level diagnosing model to classify a writer) is plausible only if the recognising model performs well while presented legible glyphs. In other words, the diagnosing model based on a poorly performing recognising model would wrongly label all children dysgraphic. Hence, we have to make sure that the recognising model performs well enough when glyphs written by non-dysgraphic children are provided as the input.

Since we are in the regime of scarcity of annotated data for dysgraphic children (as mentioned in Section \ref{scarcity}), we decided to use a straightforward diagnosing model (taking an average of given predictions) built on top of an advanced recognizing model (that may leverage the massive data we have for non-dysgraphic children).

\subsubsection{Assessing glyphs' legibility}\label{dataset_div}

We divided our data into three disjoint sets:
\begin{itemize}
\item \emph{Training set} -- $80\%$ of \emph{TD dataset}.
\item \emph{Validation set} -- $20\%$ of \emph{TD dataset}.
\item \emph{Dysgraphic set} -- consists of all glyphs written by dysgraphic children (\emph{D dataset}).
\end{itemize}

Hence, the recognising model is trained on examples which are believed to be more legible.

The model's input consists of a single glyph, and there is a class for every glyph. Depending on the model used, the inputs may be the trajectories of a pen (point-by-point for the RNN case) or the static images (final, visual results for the CNN). This means the model treats the problem as a multi-class classification balanced by the design of the data collecting procedure (each child has to write down every 36 glyphs).
\\
In real-life applications, a child would be asked to write down a given glyph and, even if the ground truth is known, we let the model predict the glyph's label. Afterwards, the discrepancy between the ground truth and the model's prediction is measured.
\\
Assuming that the model discriminates the glyphs properly with high accuracy, high discrepancy means that glyphs are not legible and the child is likely to be dysgraphic. Therefore, while each child writes down all glyphs, the model predicts in parallel the probability for every glyph to be the requested one. These glyph-level scores are averaged per child, and this value is understood as a statistic for evaluating writing proficiency, and we call this statistic $D$ (\emph{dysgraphic statistic}).

Taking an unweighted average is a simple, arbitrary choice and more sophisticated methods may be invented as can be seen in Section \ref{sec:DiscriminativeLetters}. However, we found this approach works sufficiently well.

\subsubsection{Final dysgraphic prediction}
The lower the value of \emph{D statistic}, the more likely the child will be dysgraphic. However, to solve our main problem, we have to answer the following question:
\begin{center}\textit{Having a value of \emph{D statistic} for the given child, is the writer dysgraphic?}
\end{center}

In other words, we have to find the threshold (the critical value) of \emph{D statistic} that divides the children into two groups (dysgraphic and non-dysgraphic children). In the original BHK test, children are considered dysgraphic if they obtain a score beyond two standard deviations from the normative group. In practice, it was found that 8.6\% of the population is dysgraphic \cite{charles2004echelle}.
\\
We, therefore, decided to use the same approach to compute the threshold in a way that the 8.6\% of the children who obtained the lowest values of \emph{D statistic} will be considered dysgraphic by our diagnosing model. Of course, this threshold value depends on the recognising model. The main property of this approach for fixing the threshold is that, on average, $8.6\%$ of children are labelled dysgraphic, precisely like the original BHK test.

\subsection{Discrimination procedure}

To summarise, the discrimination procedure for a given child consists of the following steps:
\begin{enumerate}
\item The child is asked to write down the 36 glyphs on the same digital tablet.
\item The recognising model, which is trained to discriminate glyphs, is evaluated for all drawings and a single value (the probability that the glyph is the one requested) for each is obtained. A higher value represents, the greater legibility of the glyph.
\item All $36$ scores are averaged giving the \emph{D statistic}.
\item The value of the \emph{D statistic} describes the child's handwriting proficiency. It is then compared with a threshold calculated in such a way that, on average, $8.6\%$ of children score below. A child with a score below this threshold is labelled dysgraphic.
\end{enumerate}

\section{Results}

As mentioned in Section \ref{sec:data}, we trained the model to calculate the \emph{D statistic} for a given child. This value is compared with the threshold found previously. If the value of \emph{D statistic} is below the threshold, then the child will be labelled "dysgraphic."
\\
In this section we address the following question:
\begin{center}\emph{Do $8.6\%$ of all children suspected by our prototype represent the group of children truly diagnosed as dysgraphic by the original BHK test?}
\end{center}

We show in this section that our model provides very promising results when it uses dynamics of writing as an input.

\subsection{Recurrent neural network}
\label{RNN}
We present results obtained using recurrent neural networks which, as pointed out in Subsection \ref{sec:neuralNetworks}, utilises the temporal aspect of the data to take into account the dynamics of writing. We argue that this temporal information provides valuable insights so that our model is advantageous over other methods only using static images (including the original BHK test).

\begin{figure}[ht]
\centering
\includegraphics[width=0.95\linewidth]{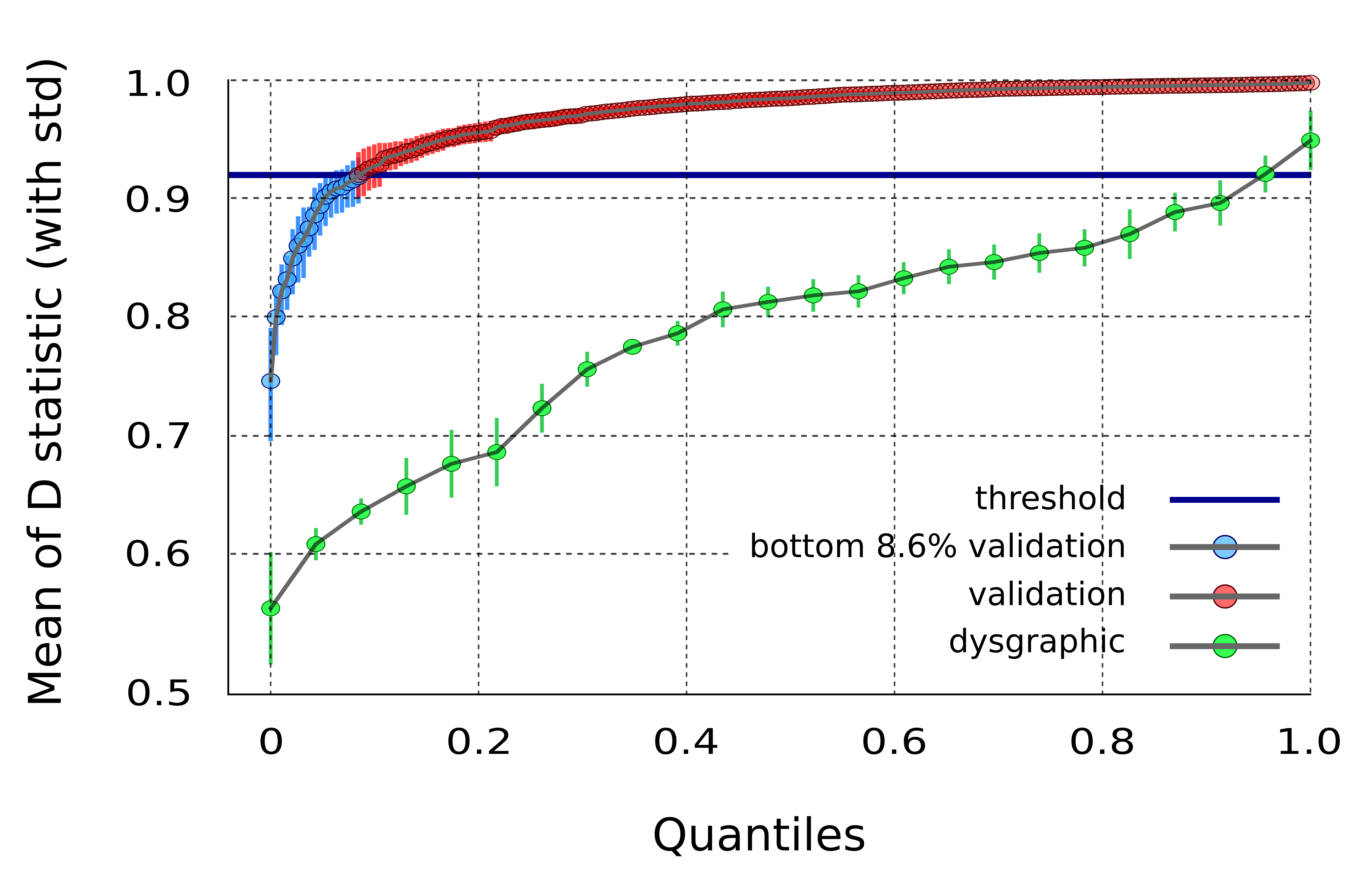}
\caption{With RNN, quantile function of averaged \emph{D statistic} on validation and dysgraphic sets. The blue horizontal line represents the estimated threshold. Each point represents the \emph{D statistic} for a given child (averaged on each fold, the error bar represents the standard deviation). By definition, $8.6\%$ of the validation results are below threshold which separates the children into dysgraphic and non-dysgraphic groups, such that in the perfect case, all dysgraphic children should score below this threshold.}
\label{fig:allRNN}
\end{figure}

A k-fold cross-validation \cite{friedman2001elements} (with $k = 5$) produced the graph presented in Figure \ref{fig:allRNN}. Details about the model, the learning procedure, and the training procedure may be found in the Appendix.
In Figure \ref{fig:allRNN}, the x-axis are quantiles and the y-axis includes averaged \emph{D statistic} values. By definition, $8.6\%$ of validation results are below the threshold, which separates the children into dysgraphic and non-dysgraphic groups. It means that in the perfect case all dysgraphic children should score below the threshold. The most important finding is that only two dysgraphic children are not below the threshold, and hence more than $90\%$ of dysgraphic children are correctly diagnosed. This means that even the prototype of the digital BHK test (trained only on single glyphs) that takes into account the dynamics of writing would identify nearly all children labelled by the original BHK test.

This result is impressive as without any further research, we can prepare a model capable of identifying more than $90\%$ of all dysgraphic children without the laborious review by a human expert. Additionally, this digital approach for labelling children is highly objective. Hence, this work is a milestone in achieving our long-term mission of creating a digital BHK test.

\subsection{Convolutional neural network}
As explained in Subsection \ref{sec:neuralNetworks}, CNN is a category of neural networks shown to be very effective in tasks, such as image recognition and classification. They work only with static 2D data, so, contrary to the recurrent neural network (see Subsection \ref{RNN}), the dynamics of handwriting (i.e., the timeframe) cannot be taken into account. Only the final trace will be used by the model to predict the label of the data. 

\begin{figure}[ht]
\centering
\includegraphics[width=0.95\linewidth]{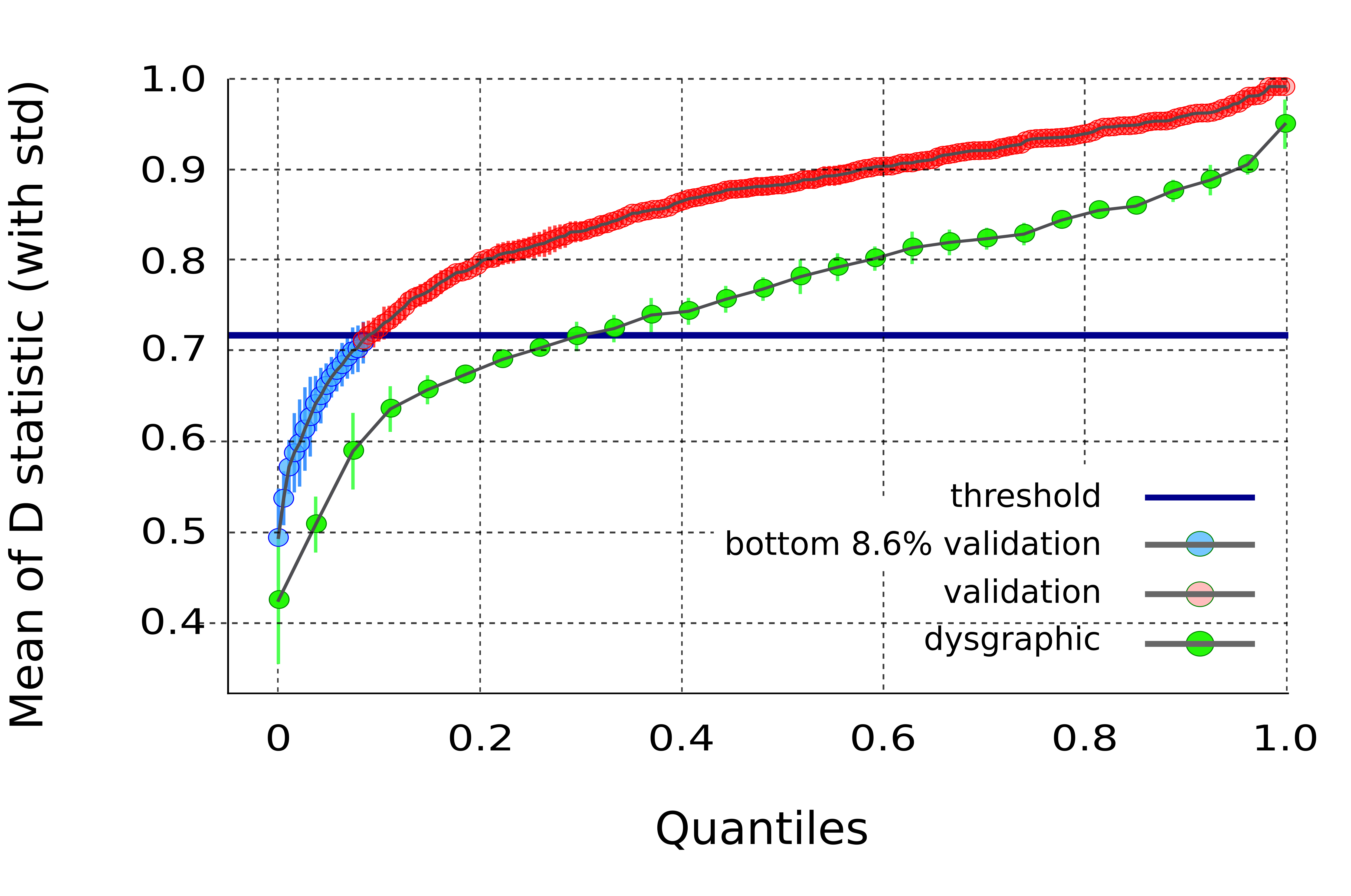}
\caption{With CNN model, quantile function of \emph{D statistic} on validation and dysgraphic sets. The estimated threshold is included, and each point represents \emph{D statistic} for one child (averaged on each fold, the error bar represents the standard deviation). By definition, $8.6\%$ of the validation results are below the threshold, which divides the children into dysgraphic and non-dysgraphic groups such that, in the perfect case, all dysgraphic children should score below the threshold. In our case, the model is not very efficient as less than $30\%$ of dysgraphic children are correctly diagnosed.}
\label{fig:allCNN}
\end{figure}

Figure \ref{fig:allCNN} follows the same procedure as used to generate Figure \ref{fig:allRNN} with the only difference being that CNN is the recognising model instead of RNN. It can be seen that even if the score of dysgraphic children is significantly lower than the one not diagnosed (which does not guarantee all these children are non-dysgraphic), we failed to obtain a clear separation of 90\% of the subjects as found with the RNN. As can be seen in Figure, only 25-30\% of the dysgraphic are below the threshold line suggesting that the model can successfully identify approximately a quarter of the dysgraphic children.

The results presented in this section suggest the superiority of the RNN over CNN for this task. We believe this difference in efficiency is due to the incorporation of the movement's dynamics for the RNN. This assumption will be further explored in Section \ref{sec:confusion}.

\subsection{Discriminative glyphs}
\label{sec:DiscriminativeLetters}
The method used to calculate the \emph{D statistic} for children may be slightly modified to obtain a similar score for glyphs. We can calculate glyph-level (not child-level) \emph{D statistic} separately for the dysgraphic children (\emph{D dataset}) and for the others (\emph{TD dataset}) (validation part only) to identify the most \emph{discriminative glyphs}. These glyphs are particularly hard to write for dysgraphic children compared to the general population making them possibly the most helpful for identifying dysgraphic children.

\begin{figure}[ht]
\centering
\includegraphics[width=0.95\linewidth]{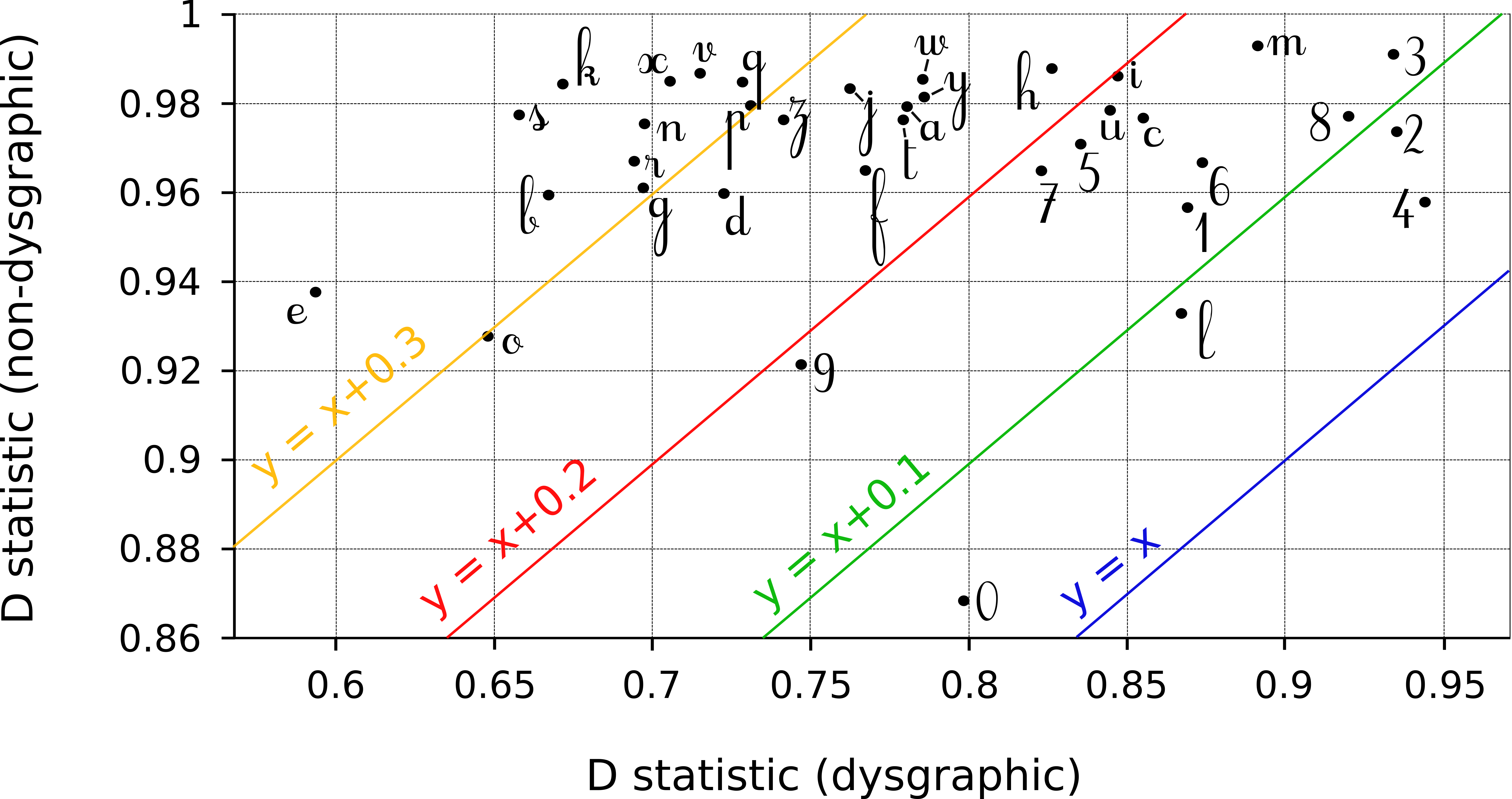}
\caption{\emph{Discriminative glyphs} (based on RNN model). The comparison of the difficulty in writing particular glyphs for dysgraphic and non-dysgraphic children. Glyphs above the yellow line are harder to draw for the group of children with writing difficulties. There are no glyphs below the blue line since each letter is easier for non-dysgraphic children. However, the digits '0' and '4' are nearly of the same difficulty for both groups (it may be initially surprising that the digit '0' is harder than the digit '4', but this is the case because children are expected to write down cursive glyphs; similarly, the letter 'o' and the digit '0' are found to be hard).}
\label{fig:lettersRNN}
\end{figure}

In Figure \ref{fig:lettersRNN}, the \emph{D statistics} (averaged for each student and averaged on the five folds, and the standard deviation were not presented for clarity) of every glyph are presented. For each, the x and y coordinates represent the mean \emph{D statistic} for the dysgraphic and non-dysgraphic children, respectively. Each glyph close to the blue line (where dysgraphic children and "average" children obtain the same score) is approximately of the same difficulty for dysgraphic and non-dysgraphic children. Those above the yellow line are harder to draw for the group of children with writing difficulties.

\begin{figure}[ht]
\centering
\includegraphics[width=0.95\linewidth]{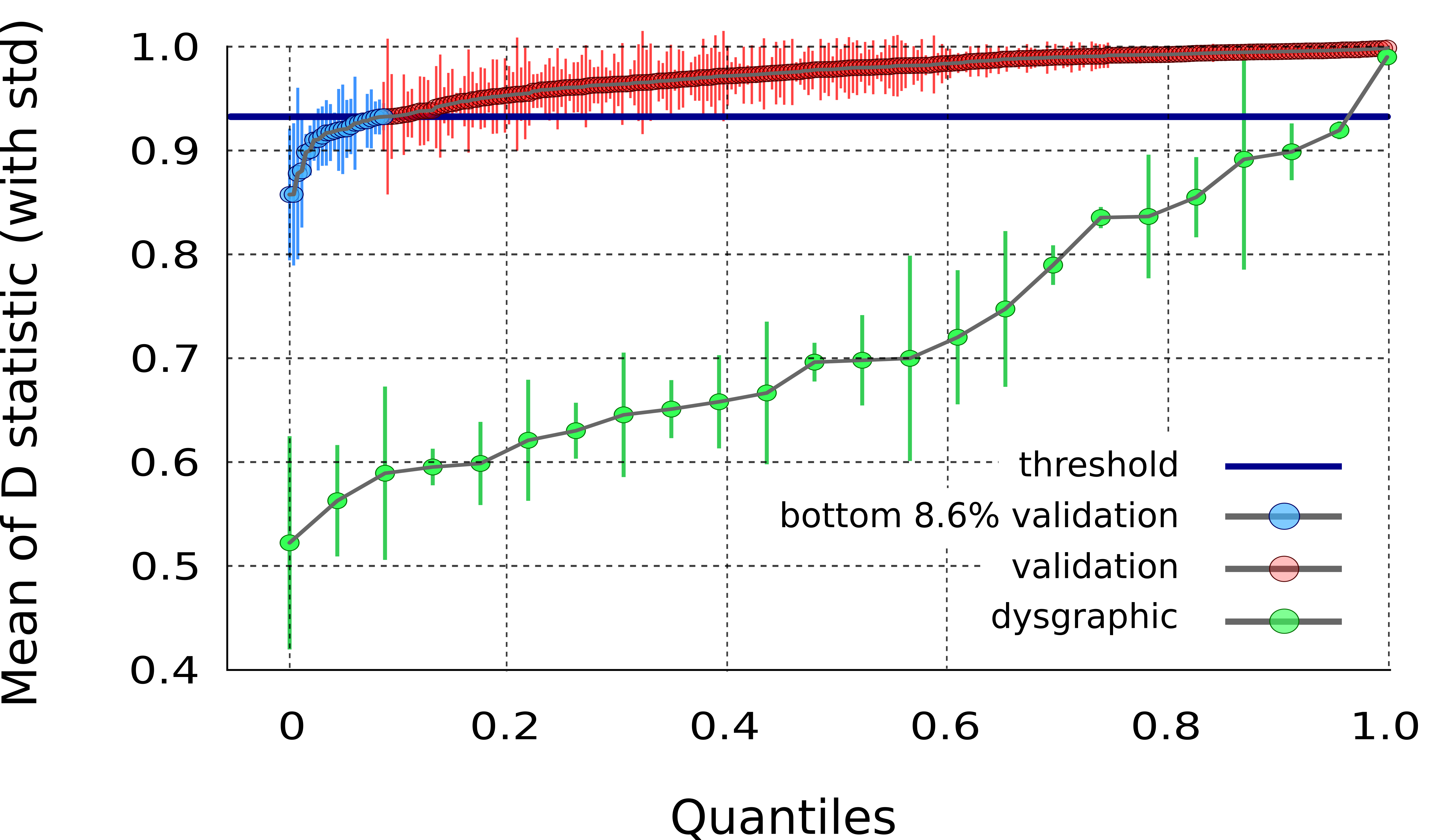}
\caption{With RNN, quantile function of \emph{D statistic} on the validation and dysgraphic sets calculated for the 15 most discriminative glyphs. The estimated threshold is also included. Each point represents \emph{D statistic} for one child (averaged on each fold, the error bar represents the standard deviation). The results are similar to the case where all glyphs are used (see Figure \ref{fig:allRNN}). The key difference is a higher variance that reflects a smaller number of the instances for each child.}
\label{fig:RNNFL}
\end{figure}

Therefore, the glyphs 'e', 's', and 'k', for example, are \emph{discriminative glyphs} according to our RNN model. This knowledge may be used to prepare even a better prototype, for example, a weighting factor may be assigned proportionally to the discriminative power of the glyphs. Another possibility would be to ask children to write down only the most discriminative glyphs. This technique was applied to create Figure \ref{fig:RNNFL}. The 15 most discriminative glyphs were used to compute the \emph{D Statistic}. This technique reduces the diagnostic test duration while providing an accuracy similar to when all glyphs are used for the diagnosis. As mentioned in Subsection \ref{RNN}, we used a cross-validation technique by training the model for five folds. For each fold, the most discriminative glyphs found can be slightly different since they are within this particular model and validation set. This means that using the same discriminative glyphs to assess the model performance on the same validation set would bias the model. Hence, to alleviate this issue, we selected 15 of the most discriminative glyphs found for another fold, where each fold was used to obtain the most discriminative glyph only once.

It is interesting to notice that some of the most discriminative letters found are not used in the original BHK ('y', 'k', 'q', 'w', and 'x'). We believe the utilisation of this subset in our prototype is one of the reasons why our model requires fewer data compared to the BHK for classification.

\subsection{Adding the handwriting dynamic reduces confusion between visually similar glyphs}
\label{sec:confusion}

\begin{figure}[ht]
\centering
\includegraphics[width=0.95\linewidth]{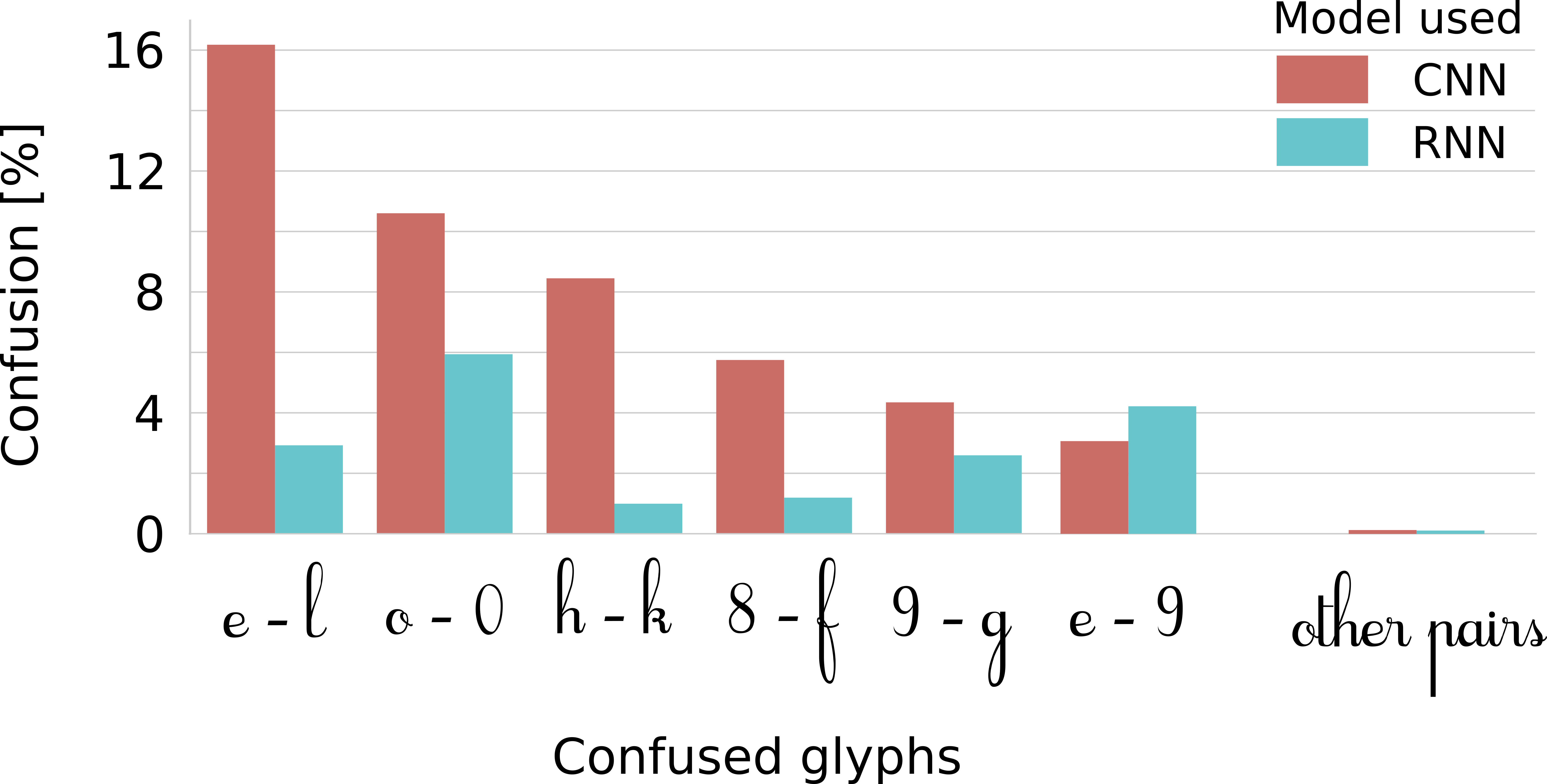}
\caption{Confusion between the six most similar pair of glyphs (those that bring the highest confusion) for the RNN and CNN models.}
\label{fig:confusion}
\end{figure}

Certain glyphs might look very similar when we only have access to the final trace. This is, for example, the case between the letters 'e' and 'l' or between 'g' and the digit '9'. Thus, we hypothesize that the final trace is sometimes not enough to discriminate between these classes, so, in these cases, the dynamics of handwriting must be fundamental. For these glyphs whose final traces look similar, the CNN will make inaccurate classifications while the RNN will have better results due to its access to the movement dynamics. We believe that this is one of the reasons explaining the superiority of the RNN model over CNN. Figure \ref{fig:confusion} plots the confusion between the six most similar pairs of glyphs (representing the pairs of glyphs with the greatest confusion) of the CNN and RNN models. 
The confusion for a pair of glyph A-B is the number of misclassification (B instead of A or A instead of B) over the entire set of As and Bs. The misclassification of A in B over all As and the misclassification of B in A over all Bs are different. However, as it was giving similar values in our case, we decided to use their average to obtain the final misclassification values plotted in Figure \ref{fig:confusion}. For example, a confusion of 16\% for the CNN model is seen between the letters 'e' and 'l' meaning that the misclassification of 'e' in 'l' or the misclassification of 'l' in 'e' occurs in 16\% of the cases.
\\
This graph appears to confirm our hypothesis as we can see that the RNN generates fewer confusions between the pairs of glyphs that looks \emph{visually} similar, likely due to its access to the handwriting dynamics. Concerning the other pairs of glyphs that do not look \emph{visually} similar, the confusion between the two models is essentially the same and close to 0\% of misclassification.

\subsection{Additional class}

We noticed that it is beneficial to add a class to the training set, which we call the star or, simply, the '*' class. This class consists of random hybrids of real glyphs obtained by combining two or three drawings to create a non-existing glyph, such that the beginning part of the first glyph trajectory is associated to the middle part of the second one, for example.

This data augmentation makes our model suspicious about glyphs that look strange and labels them as '*'. In other words, our model cannot assume that the analysed object is a glyph. We believe this strategy helps in situations when a dysgraphic child would perform poorly on a glyph but still makes a good guess since the mistake is easy to figure out. Examples include triple 'w' or a letter 'i' with additional dots.

\begin{figure}[ht]
\centering
\includegraphics[width=0.95\linewidth]{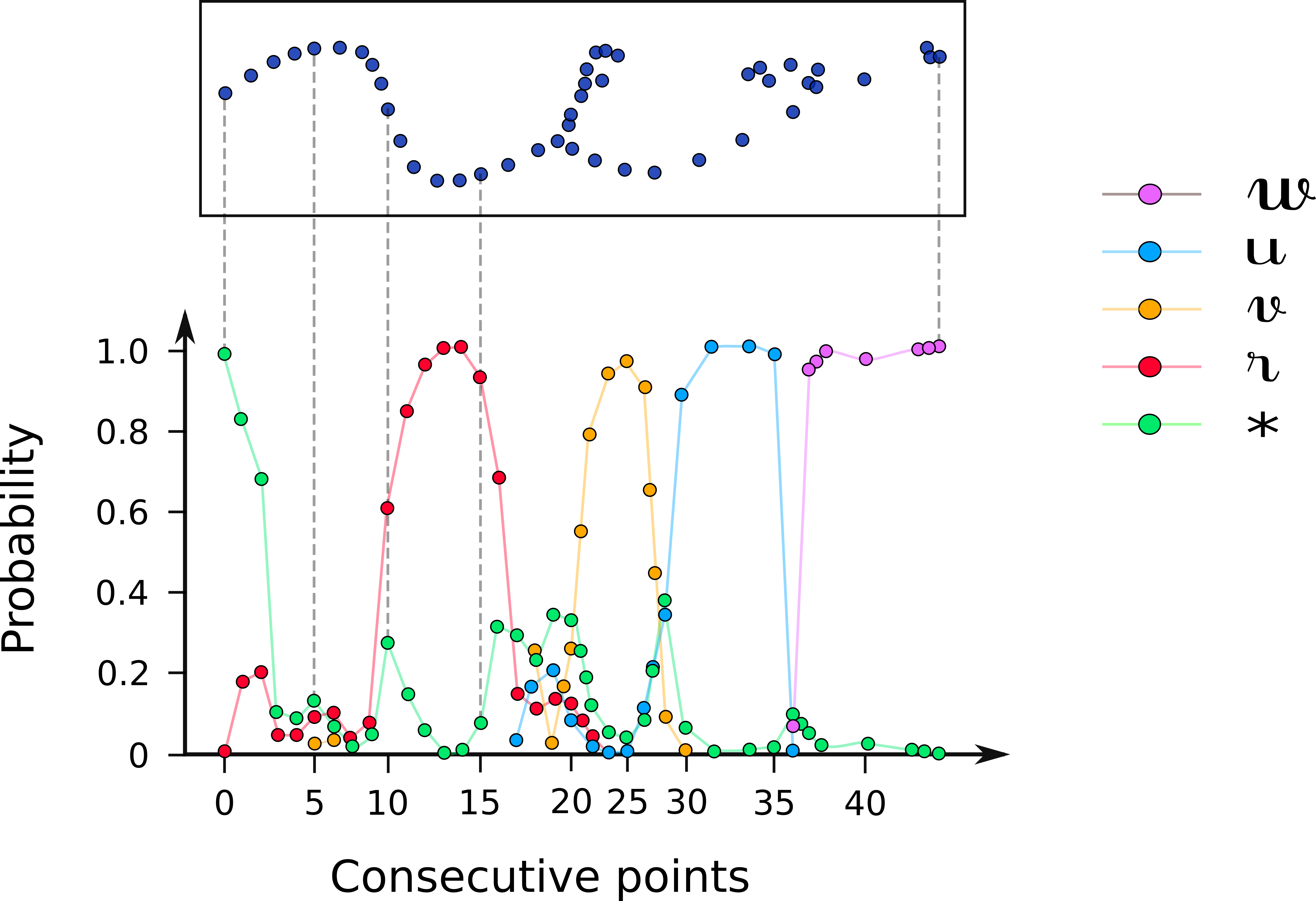}
\caption{An example of glyph 'w' (top) with consecutive RNN label predictions (bottom).}
\label{fig:w_d}
\end{figure}

Figure \ref{fig:w_d} presents the evolution of consecutive predictions of our RNN model for the letter 'w'. In the beginning, the most probable answers are star class since the usual existing glyphs are always longer. Afterward, the model switches to 'r' because the first curve of the glyph 'w' looks similar to the beginning of the glyph 'r'. The model further evolves to 'v', 'u', and finally the correct answer 'w'. Employing the dynamics in the model not only improves the predictive power but also makes the model interpretable.



\section{Discussion}
\label{Conclusion}

This tablet-based test might pave the way to the systematic diagnosis of dysgraphia in schools. Our aim would be to minimize the impact of dysgraphia on children’s school experiences by providing a rapid, simple and cheap diagnostic tool that will allow for earlier remediation. For example, this tool could be employed in conjunction with other screening tests that are already in use in schools, such as auditory and ophthalmologic tests, facilitating the early detection and remediation of handwriting issues.

Contrary to all the techniques used for the diagnosis of dysgraphia (particularly the BHK test) that only include the final trace as input and, therefore, exclude key parts of the information, our test exploits the dynamic of the handwriting. We showed that adding the dynamics of the movement improves the accuracy and decreases the necessary amount of information needed to deliver the identification of dysgraphic children. The RNN using the time-frame in addition to the trace allows improved results (90\% of dysgraphic children successfully labelled dysgraphic) over the CNN using only the static image (less than 30\% of dysgraphic children successfully labelled dysgraphic). This is the primary reason that suggests developing a digital approach is valuable for this field as the dynamics of the movement is difficult to interpret with the final trace alone, even if an expert is observing the child while writing.

A digital approach to the diagnosis offers additional potential interests. First, it enables to remove human subjectivity as the model cannot be biased by any external parameters that can influence a human. The diagnosis of dysgraphia is important as it has significant implications for the lives of children and their parents. Hence, providing a non-subjective method is certainly not negligible. Moreover, digitally diagnosing dysgraphia will enable to save time as trained therapists spend around 10 minutes per BHK test to analyse the handwriting of the children.

Other information, such as the pressure or inclination of the pen during handwriting extractable with devices like the Wacom(C) graphics tablet could be integrated with the model to improve efficiency further. Currently, the children are asked to write all the letters of the alphabet. As seen in Section \ref{sec:DiscriminativeLetters}, even if we could remove certain nondiscriminatory letters, some vital information may remain in the way letters or words are linked together. Our model does not consider this potential connection, and it would be interesting to incorporate this aspect in future work.

During this project, the data acquisition utilised a Wacom(C) tablet with an A5 paper sheet affixed to the screen. This device was selected because, in addition to the x, y, and time coordinates, it can accurately extract pressure and pen inclination. However, our current model only incorporates the first three properties. Data acquisition with a high-end tablet, such as the one we use, may soon become obsolete as an available Android device that is less expensive and more accessible may offer a practical solution for anyone wishing to apply the model. However, it is important to note that the friction between the pen and a tactile surface or paper is not the same between the two classes of device. So, data extracted from the two solutions can differ in ways that would mislead the model. In future work, it may be required to compare the handwriting output difference between the two approaches and, if necessary, train a new model with data extracted from a tactile tablet.

\section{Future Work}
In a general context, this approach could have great potential for educational purposes, such as the CoWriter project \cite{hood2015children,jacq2016building,johal2016child,lemaignan2016learning}. This project aims to teach handwriting by using an original approach: the learner becomes the professor of a robot needing help to improve writing techniques. This paradigm, known as “learning by teaching,” takes advantage of the Prot\'{e}g\'{e} effect by making the child feel responsible for the robot, thus encouraging the child to be more engaged in the task. Words the child needs to teach to the robot, using a digital tablet to write handwriting examples, are randomly chosen and might not be appropriate for the child's skill level or needs. We imagine applying this work to create a child's handwriting profile using the D-statistic of each letter to represent the child's proficiency for the letter. With this type of handwriting profile, we can then choose more relevant words that would make the child focus on the targeted letters the model detects with deficiencies.

\section*{Acknowledgments}

We would like to thank the Swiss National Science Foundation for supporting this project through the Swiss National Center of Competence in Research Robotics, the French National Center for Scientific Research (CNRS), and the University Grenoble-Alpes (France). Konrad \.Zo\l{}na would also like to thank the National Science Centre, Poland (\mbox{2017/27/N/ST6/00828}, \mbox{2018/28/T/ST6/00211}). We are all grateful to the directors and teachers of the participating schools, to the Reference Center for Learning and Speech Disorders at the Grenoble hospital, and to all children and their parents for participating in the study.

\section*{Statements of ethical approval}
The present study was conducted in accordance with the Declaration of Helsinki, was approved by the University ethics committee (agreement n° 2016-01-05-79). It was conducted with the understanding and written consent of each child's parents and the oral consent of each child and in accordance with the ethics convention between the academic organisation (LPNC-CNRS) and educational organisations

\bibliographystyle{abbrv}
\bibliography{sample}

\begin{thebibliography}{10}

\bibitem{accardo2013development}
A.~P. Accardo, M.~Genna, and M.~Borean.
\newblock Development, maturation and learning influence on handwriting
  kinematics.
\newblock {\em Human movement science}, 32(1):136--146, 2013.

\bibitem{alamargot2015does}
D.~Alamargot and M.-F. Morin.
\newblock Does handwriting on a tablet screen affect students’ graphomotor
  execution? a comparison between grades two and nine.
\newblock {\em Human movement science}, 44:32--41, 2015.

\bibitem{amundson1995evaluation}
S.~J. Amundson.
\newblock {\em Evaluation Tool of Children's Handwriting: ETCH examiner's
  manual}.
\newblock OT KIDS, 1995.

\bibitem{asselborn2018automated}
T.~Asselborn, T.~Gargot, {\L}.~Kidzi{\'n}ski, W.~Johal, D.~Cohen, C.~Jolly, and
  P.~Dillenbourg.
\newblock Automated human-level diagnosis of dysgraphia using a consumer
  tablet.
\newblock {\em npj Digital Medicine}, 1(1):42, 2018.

\bibitem{asselborn2018bringing}
T.~Asselborn, A.~G{\"u}neysu~{\"O}zg{\"u}r, K.~Mrini, E.~Yadollahi,
  A.~{\"O}zg{\"u}r, W.~Johal, and P.~Dillenbourg.
\newblock Bringing letters to life: handwriting with haptic-enabled tangible
  robots.
\newblock In {\em Proceedings of the 17th ACM Conference on Interaction Design
  and Children}, number CONF, pages 219--230. ACM, 2018.

\bibitem{ayres1912scale}
L.~P. Ayres.
\newblock {\em A scale for measuring the quality of handwriting of school
  children}.
\newblock Number 113. Russell Sage Foundation. Dept. of Child Hygiene, 1912.

\bibitem{berninger1998language}
V.~W. Berninger and S.~Graham.
\newblock Language by hand: A synthesis of a decade of research on handwriting.
\newblock {\em Handwriting Review}, 12(1):11--25, 1998.

\bibitem{berninger1991scientific}
V.~W. Berninger, D.~T. Mizokawa, and R.~Bragg.
\newblock Scientific practitioner: Theory-based diagnosis and remediation of
  writing disabilities.
\newblock {\em Journal of school psychology}, 29(1):57--79, 1991.

\bibitem{bezzi1962standardized}
R.~Bezzi.
\newblock A standardized manuscript scale for grades 1, 2, and 3.
\newblock {\em The Journal of Educational Research}, 55(7):339--340, 1962.

\bibitem{blote1991longitudinal}
A.~W. Bl{\"o}te and L.~Hamstra-Bletz.
\newblock A longitudinal study on the structure of handwriting.
\newblock {\em Perceptual and Motor Skills}, 72(3):983--994, 1991.

\bibitem{bourdin2000graphic}
B.~Bourdin and M.~Fayol.
\newblock Is graphic activity cognitively costly? a developmental approach.
\newblock {\em Reading and Writing}, 13(3):183--196, 2000.

\bibitem{charles2004echelle}
M.~Charles, R.~Soppelsa, and J.~Albaret.
\newblock {\'E}chelle d’{\'e}valuation rapide de l’{\'e}criture chez
  l’enfant (bhk).
\newblock {\em Paris: {\'E}dition du Centre de psychologie appliqu{\'e}e},
  2004.

\bibitem{chartrel2008impact}
E.~Chartrel and A.~Vinter.
\newblock The impact of spatio-temporal constraints on cursive letter
  handwriting in children.
\newblock {\em Learning and instruction}, 18(6):537--547, 2008.

\bibitem{cho2014learning}
K.~Cho, B.~Van~Merri{\"e}nboer, C.~Gulcehre, D.~Bahdanau, F.~Bougares,
  H.~Schwenk, and Y.~Bengio.
\newblock Learning phrase representations using rnn encoder-decoder for
  statistical machine translation.
\newblock {\em arXiv preprint arXiv:1406.1078}, 2014.

\bibitem{christensen2009critical}
C.~A. Christensen.
\newblock The critical role handwriting plays in the ability to produce
  high-quality written text.
\newblock {\em The SAGE handbook of writing development}, pages 284--299, 2009.

\bibitem{ciregan2012multi}
D.~Ciregan, U.~Meier, and J.~Schmidhuber.
\newblock Multi-column deep neural networks for image classification.
\newblock In {\em Computer Vision and Pattern Recognition (CVPR), 2012 IEEE
  Conference on}, pages 3642--3649. IEEE, 2012.

\bibitem{danna2013signal}
J.~Danna, V.~Paz-Villagr{\'a}n, and J.-L. Velay.
\newblock Signal-to-noise velocity peaks difference: A new method for
  evaluating the handwriting movement fluency in children with dysgraphia.
\newblock {\em Research in developmental disabilities}, 34(12):4375--4384,
  2013.

\bibitem{decoste2002training}
D.~Decoste and B.~Sch{\"o}lkopf.
\newblock Training invariant support vector machines.
\newblock {\em Machine learning}, 46(1):161--190, 2002.

\bibitem{di2008dynamic}
C.~Di~Brina, R.~Niels, A.~Overvelde, G.~Levi, and W.~Hulstijn.
\newblock Dynamic time warping: A new method in the study of poor handwriting.
\newblock {\em Human movement science}, 27(2):242--255, 2008.

\bibitem{erez1999hebrew}
N.~Erez and S.~Parush.
\newblock The hebrew handwriting evaluation.
\newblock {\em School of Occupational Therapy. Faculty of Medicine. Hebrew
  University of Jerusalem, Israel}, 1999.

\bibitem{feder2007handwriting}
K.~P. Feder and A.~Majnemer.
\newblock Handwriting development, competency, and intervention.
\newblock {\em Developmental Medicine \& Child Neurology}, 49(4):312--317,
  2007.

\bibitem{freeman1959new}
F.~N. Freeman.
\newblock A new handwriting scale.
\newblock {\em The Elementary School Journal}, 59(4):218--221, 1959.

\bibitem{friedman2001elements}
J.~Friedman, T.~Hastie, and R.~Tibshirani.
\newblock {\em The elements of statistical learning}, volume~1.
\newblock Springer series in statistics New York, 2001.

\bibitem{graham2001manuscript}
S.~Graham, N.~Weintraub, and V.~Berninger.
\newblock Which manuscript letters do primary grade children write legibly?
\newblock {\em Journal of Educational Psychology}, 93(3):488, 2001.

\bibitem{graves2013generating}
A.~Graves.
\newblock Generating sequences with recurrent neural networks.
\newblock {\em arXiv preprint arXiv:1308.0850}, 2013.

\bibitem{graves2008unconstrained}
A.~Graves, M.~Liwicki, H.~Bunke, J.~Schmidhuber, and S.~Fern{\'a}ndez.
\newblock Unconstrained on-line handwriting recognition with recurrent neural
  networks.
\newblock In {\em Advances in Neural Information Processing Systems}, pages
  577--584, 2008.

\bibitem{graves2009novel}
A.~Graves, M.~Liwicki, S.~Fern{\'a}ndez, R.~Bertolami, H.~Bunke, and
  J.~Schmidhuber.
\newblock A novel connectionist system for unconstrained handwriting
  recognition.
\newblock {\em IEEE transactions on pattern analysis and machine intelligence},
  31(5):855--868, 2009.

\bibitem{hamstra1993longitudinal}
L.~Hamstra-Bletz and A.~W. Bl{\"o}te.
\newblock A longitudinal study on dysgraphic handwriting in primary school.
\newblock {\em Journal of Learning Disabilities}, 26(10):689--699, 1993.

\bibitem{hamstra1987concise}
L.~Hamstra-Bletz, J.~DeBie, and B.~Den~Brinker.
\newblock Concise evaluation scale for children’s handwriting.
\newblock {\em Lisse: Swets}, 1, 1987.

\bibitem{herrick1963evaluation}
V.~E. Herrick and A.~Erlebacher.
\newblock The evaluation of legibility in handwriting.
\newblock {\em New horizons for research in handwriting}, pages 207--236, 1963.

\bibitem{hochreiter1997long}
S.~Hochreiter and J.~Schmidhuber.
\newblock Long short-term memory.
\newblock {\em Neural computation}, 9(8):1735--1780, 1997.

\bibitem{hood2015children}
D.~Hood, S.~Lemaignan, and P.~Dillenbourg.
\newblock When children teach a robot to write: An autonomous teachable
  humanoid which uses simulated handwriting.
\newblock In {\em Proceedings of the Tenth Annual ACM/IEEE International
  Conference on Human-Robot Interaction}, pages 83--90. ACM, 2015.

\bibitem{jacq2016building}
A.~Jacq, S.~Lemaignan, F.~Garcia, P.~Dillenbourg, and A.~Paiva.
\newblock Building successful long child-robot interactions in a learning
  context.
\newblock In {\em Human-Robot Interaction (HRI), 2016 11th ACM/IEEE
  International Conference on}, pages 239--246. IEEE, 2016.

\bibitem{jaeger2001online}
S.~Jaeger, S.~Manke, J.~Reichert, and A.~Waibel.
\newblock Online handwriting recognition: the npen++ recognizer.
\newblock {\em International Journal on Document Analysis and Recognition},
  3(3):169--180, 2001.

\bibitem{johal2016child}
W.~Johal, A.~Jacq, A.~Paiva, and P.~Dillenbourg.
\newblock Child-robot spatial arrangement in a learning by teaching activity.
\newblock In {\em Robot and Human Interactive Communication (RO-MAN), 2016 25th
  IEEE International Symposium on}, pages 533--538. Ieee, 2016.

\bibitem{jolly2014analysis}
C.~Jolly and E.~Gentaz.
\newblock Analysis of cursive letters, syllables, and words handwriting in a
  french second-grade child with developmental coordination disorder and
  comparison with typically developing children.
\newblock {\em Frontiers in psychology}, 4:1022, 2014.

\bibitem{jolly2014one}
C.~Jolly, C.~Huron, and E.~Gentaz.
\newblock A one-year survey of cursive letter handwriting in a french
  second-grade child with developmental coordination disorder.
\newblock {\em L’Ann{\'e}e Psychol}, 2014.

\bibitem{karlsdottir2002problems}
R.~Karlsdottir and T.~Stefansson.
\newblock Problems in developing functional handwriting.
\newblock {\em Perceptual and motor skills}, 94(2):623--662, 2002.

\bibitem{kim2014convolutional}
Y.~Kim.
\newblock Convolutional neural networks for sentence classification.
\newblock {\em arXiv preprint arXiv:1408.5882}, 2014.

\bibitem{kingma14}
D.~P. Kingma and J.~Ba.
\newblock Adam: {A} method for stochastic optimization.
\newblock {\em CoRR}, abs/1412.6980, 2014.

\bibitem{krizhevsky2012imagenet}
A.~Krizhevsky, I.~Sutskever, and G.~E. Hinton.
\newblock Imagenet classification with deep convolutional neural networks.
\newblock In {\em Advances in neural information processing systems}, pages
  1097--1105, 2012.

\bibitem{lecun1998gradient}
Y.~LeCun, L.~Bottou, Y.~Bengio, and P.~Haffner.
\newblock Gradient-based learning applied to document recognition.
\newblock {\em Proceedings of the IEEE}, 86(11):2278--2324, 1998.

\bibitem{lemaignan2016learning}
S.~Lemaignan, A.~Jacq, D.~Hood, F.~Garcia, A.~Paiva, and P.~Dillenbourg.
\newblock Learning by teaching a robot: The case of handwriting.
\newblock {\em IEEE Robotics \& Automation Magazine}, 23(2):56--66, 2016.

\bibitem{li2010automatic}
T.~L. Li, A.~B. Chan, and A.~Chun.
\newblock Automatic musical pattern feature extraction using convolutional
  neural network.
\newblock In {\em Proc. Int. Conf. Data Mining and Applications}, 2010.

\bibitem{marti2001using}
U.-V. Marti and H.~Bunke.
\newblock Using a statistical language model to improve the performance of an
  hmm-based cursive handwriting recognition system.
\newblock {\em International journal of Pattern Recognition and Artificial
  intelligence}, 15(01):65--90, 2001.

\bibitem{mccutchen2011novice}
D.~McCutchen.
\newblock From novice to expert: Implications of language skills and
  writing-relevant knowledge for memory during the development of writing
  skill.
\newblock {\em Journal of Writing Research}, 3(1):51--68, 2011.

\bibitem{mnih2013playing}
V.~Mnih, K.~Kavukcuoglu, D.~Silver, A.~Graves, I.~Antonoglou, D.~Wierstra, and
  M.~Riedmiller.
\newblock Playing atari with deep reinforcement learning.
\newblock {\em arXiv preprint arXiv:1312.5602}, 2013.

\bibitem{overvelde2011handwriting}
A.~Overvelde and W.~Hulstijn.
\newblock Handwriting development in grade 2 and grade 3 primary school
  children with normal, at risk, or dysgraphic characteristics.
\newblock {\em Research in developmental disabilities}, 32(2):540--548, 2011.

\bibitem{piasta2010developing}
S.~B. Piasta and R.~K. Wagner.
\newblock Developing early literacy skills: A meta-analysis of alphabet
  learning and instruction.
\newblock {\em Reading research quarterly}, 45(1):8--38, 2010.

\bibitem{plamondon2000online}
R.~Plamondon and S.~N. Srihari.
\newblock Online and off-line handwriting recognition: a comprehensive survey.
\newblock {\em IEEE Transactions on pattern analysis and machine intelligence},
  22(1):63--84, 2000.

\bibitem{pontart2013influence}
V.~Pontart, C.~Bidet-Ildei, E.~Lambert, P.~Morisset, L.~Flouret, and
  D.~Alamargot.
\newblock Influence of handwriting skills during spelling in primary and lower
  secondary grades.
\newblock {\em Frontiers in psychology}, 4:818, 2013.

\bibitem{ritchey2008building}
K.~D. Ritchey.
\newblock The building blocks of writing: Learning to write letters and spell
  words.
\newblock {\em Reading and writing}, 21(1-2):27--47, 2008.

\bibitem{rosenblum2005using}
S.~Rosenblum.
\newblock Using the alphabet task to differentiate between proficient and
  nonproficient handwriters.
\newblock {\em Perceptual and motor skills}, 100(3):629--639, 2005.

\bibitem{rosenblum2017identifying}
S.~Rosenblum and G.~Dror.
\newblock Identifying developmental dysgraphia characteristics utilizing
  handwriting classification methods.
\newblock {\em IEEE Transactions on Human-Machine Systems}, 47(2):293--298,
  2017.

\bibitem{rosenblum2003product}
S.~Rosenblum, P.~L. Weiss, and S.~Parush.
\newblock Product and process evaluation of handwriting difficulties.
\newblock {\em Educational Psychology Review}, 15(1):41--81, 2003.

\bibitem{rueckriegel2008influence}
S.~M. Rueckriegel, F.~Blankenburg, R.~Burghardt, S.~Ehrlich, G.~Henze,
  R.~Mergl, and P.~H. Driever.
\newblock Influence of age and movement complexity on kinematic hand movement
  parameters in childhood and adolescence.
\newblock {\em International Journal of Developmental Neuroscience},
  26(7):655--663, 2008.

\bibitem{simard2003best}
P.~Y. Simard, D.~Steinkraus, J.~C. Platt, et~al.
\newblock Best practices for convolutional neural networks applied to visual
  document analysis.
\newblock In {\em ICDAR}, volume~3, pages 958--962. Citeseer, 2003.

\bibitem{soppelsa2012evaluation}
R.~Soppelsa and J.~Albaret.
\newblock Evaluation de l’{\'e}criture chez l’adolescent. le bhk ado.
\newblock {\em Entretiens de Psychomotricit{\'e}}, pages 66--76, 2012.

\bibitem{thorndike1912handwriting}
E.~L. Thorndike.
\newblock {\em Handwriting}.
\newblock Teachers College, Columbia University, 1912.

\bibitem{torrey2009transfer}
L.~Torrey and J.~Shavlik.
\newblock Transfer learning.
\newblock {\em Handbook of Research on Machine Learning Applications and
  Trends: Algorithms, Methods, and Techniques}, 1:242, 2009.

\bibitem{vinter2010effects}
A.~Vinter and E.~Chartrel.
\newblock Effects of different types of learning on handwriting movements in
  young children.
\newblock {\em Learning and Instruction}, 20(6):476--486, 2010.

\bibitem{wang2012end}
T.~Wang, D.~J. Wu, A.~Coates, and A.~Y. Ng.
\newblock End-to-end text recognition with convolutional neural networks.
\newblock In {\em Pattern Recognition (ICPR), 2012 21st International
  Conference on}, pages 3304--3308. IEEE, 2012.

\bibitem{ziviani2006development}
J.~Ziviani and M.~Wallen.
\newblock The development of graphomotor skills.
\newblock In {\em Hand function in the child: Foundations for remediation},
  volume 2nd edition, pages 217--236. Mosby, St-Louis, MO, 2006.

\end{thebibliography}

\clearpage
\section*{Appendix}

\subsection*{Recurrent neural network details}\label{rnnDetails}

\subsubsection*{Architecture}

The recurrent neural network is a 2-layer LSTM architecture with hidden sizes of 100 (for both layers). The output of the second LSTM layer is followed by a feed-forward neural network with one hidden layer (of size 40), a dropout (p=0.5), and ReLU (rectified linear unit) nonlinearity to obtain a final output of size 37. The final output is then softmaxed to achieve a probability distribution over all possible glyphs (26 letters, 10 digits, and the special character '*').

\subsubsection*{Learning procedure}

All parameters were initialised randomly using a uniform distribution on the interval ($-0.08$, $0.08$). The Adam optimizer \cite{kingma14} was used with default hyper-parameters except for the learning rate where $0.005$ was used instead. Gradient clipping was also used (threshold equals $10$ for $L^{\infty}$ norm), and the batch size was constant at $20$. The learning procedure was stopped after $15$ epochs without progress on the validation set (considered to be early stopping). The hyper-parameters were found using a random grid search, and the span of the checked values was not very wide due to limited computation power so that these initialization results may be improved.

\subsection*{Cross-validation}

A $k$-fold ($k=5$) cross-validation was performed for the training. For each fold, the ratio between the training and validation data was fixed to $80\% - 20\%$ (see Subsection \ref{dataset_div} for more details). Since $k$ was set to $5$ and $20\%$ of the data were used for the validation in each run, every child record in our database was used in the validation set exactly once.

\end{document}